\documentclass{PoS}

\usepackage[dvips]{graphicx}
\usepackage{float}
\usepackage{amsmath}
\usepackage{amssymb}
\usepackage{amsfonts}
\usepackage{array}

\usepackage{solscale}

\PoS{PoS(LAT2005)301}

\title{Topologically non-trivial field configurations - interplay of vortices and Dirac eigenmodes }

\author{\speaker{Stefan Solbrig}$^a$\thanks{supported by BMBF and DFG},
        \mbox{Jochen Gattnar}$^b$, \mbox{Christof Gattringer}$^c$,
        \mbox{Kurt Langfeld}$^b$, 
	\mbox{Hugo Reinhardt}$^b$\thanks{supported in part by DFG Re 856/5-1}, 
        \mbox{Andreas Sch\"afer}$^a$, \mbox{Torsten Tok}$^b$\\
        \llap{$^a$}University of Regensburg, Germany\\
        \llap{$^b$}University of T\"ubingen, Germany\\
        \llap{$^c$}University of Graz, Austria\\
        E-mail: \email{stefan.solbrig@physik.uni-r.de}}

\abstract{
  Gluon field configurations with non-trivial topology like instantons,
  magnetic monopoles and center vortices play a crucial role
  in QCD and, in particular, for the spontaneous breaking of chiral
  symmetry. Moreover,
  center vortices are strongly correlated with confinement. We present
  evidence, that there is a deep connection between the topology of gauge
  fields and center vortices.
  We use the chirally improved lattice Dirac operator to compute
  eigenvectors and eigenvalues of various lattice gauge field configurations.
  Removing vortices from thermalized configurations also removes the
  topological content of the gauge field. As a consistency check, we apply
  random changes to the raw configurations.
}

\FullConference{XXIIIrd International Symposium on Lattice Field Theory\\
                 25-30 July 2005\\
                 Trinity College, Dublin, Ireland}

\ShortTitle{Vortices \& Dirac Eigenmodes}

\begin{document}
\bibliographystyle{posutphys}

\section{Introduction}

It has long been known that the low lying eigenmodes of the 
Dirac operator are a good tool to detect infrared, localized 
structures in lattice gauge fields. In addition, the spectrum of the 
Dirac operator allows to detect chiral symmetry breaking 
via the Banks-Casher relation~\cite{Banks:docm}. The breaking of chiral 
symmetry is usually attributed to instanton-like structures, while center
vortices are considered to be responsible for confinement. However, 
both structures, instantons and vortices, give rise to zero modes of 
the Dirac operator \cite{reinhardtmodes:docm}. This is in accord with the
Atiyah-Singer-Index theorem~\cite{singer:docm}, since both types of
configurations carry topological charge. In the case of center vortices
topological charge is realized by vortex 
intersection~\cite{Engelhardt:1999xw,Reinhardt:2001kf} and
writhing \cite{Reinhardt:2001kf}.

We will use the properties of Dirac eigenvalues and eigenvectors 
to investigate the relation between vortices, chiral symmetry and 
topology~\cite{Gattnar:2004gx}. 
We used quenched \mbox{$SU(2)$} configurations. 
All the results presented here are for the 
Chirally Improved Dirac operator~\cite{Gattringer:2000js,Gattringer:2000qu}.

The first step is to identify vortices on the lattice. 
This is usually done after a suitable gauge fixing. A common choice 
is the maximal center gauge. In our case, the quantity 
\begin{gather}
  \label{eq:gaugefixing}
  R=\frac{1}{4V}\sum_{x\in\text{lattice}}\sum_\mu
  \left[ 
    \quad
    \frac{1}{2}\text{ tr }
    \left( U_\mu^g(x) \right)
    \quad
  \right]
  \left[ 
    \quad 
    \overline{ 
      \frac{1}{2}\text{ tr } %
      \left( U_\mu^g(x) \right)  
    }
    \quad
  \right]
\end{gather}
is maximized by applying gauge transformations $g$. 
In the above equation, $V$ is the number of lattice sites and 
\mbox{$U_\mu^g(x)$} is the link variable after a gauge transformation $g$. 

For \mbox{$SU(2)$}, the center projected link variable 
\mbox{$Z_\mu(x)$} is defined as: 
\begin{gather}
  \label{eq:centerprojection}
  Z_\mu(x)=\text{ sign Re tr }\left( U_\mu(x) \right )\;. 
\end{gather}

\section{Vortex Removal}

If we want to understand how vortices influence the spectrum of the 
Dirac operator, the following idea might come to your mind: 
Perform gauge fixing and a subsequent center projection as described 
above, in order to produce ``vortex only'' configurations. 
Then compute the eigenvalues and eigenvectors for these configurations 
directly. 
However, it turns out that this approach is too naive. The center 
projected configurations consist entirely of links that are either 
$+\mathbf{1}$ or $-\mathbf{1}$. In this sense, they are maximally 
discontinuous and not suitable for the analysis with our Dirac operator. 
Another approach has to be followed. 

The method of choice is the removal of center vortices, as 
introduced in \cite{deForcrand:1999ms}. The new configurations, 
called ``vortex removed configurations,'' are given by 
\begin{gather}
  \label{eq:vortexremoved}
  U_\mu^{\text{rem.}}(x)=U_\mu(x)\;Z_\mu^\dagger(x)
\end{gather}
for every lattice site $x$. They are void of center vortices 
in the sense that center projection of a vortex removed configuration 
yields a trivial gauge configuration. 

\begin{figure}[t!]
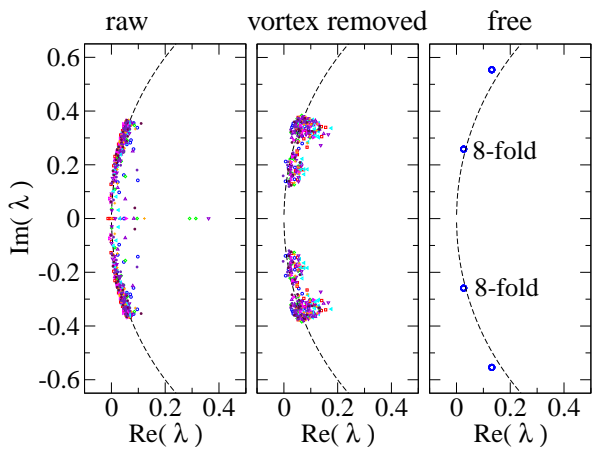

  \begin{center}
    \begin{tabular}{@{}c@{}c@{}c@{}}
      raw & vortex removed & free \\
      \includegraphics[clip=true,scale=\solGetCommonScale]
      {eigen_plot_B250F04ph05_normal_forconf}&
      \includegraphics[clip=true,scale=\solGetCommonScale]
      {eigen_plot_B250F04ph05_removed_forconf}&
      \includegraphics[clip=true,scale=\solGetCommonScale]
      {eigen_plot_B250F04ph05_free_forconf} 
    \end{tabular}
  \end{center}
  \caption[Dirac spectra]
  { Dirac spectra for the Chirally Improved Dirac
    operator for various gauge field configurations. Each plot contains 
    the superimposed spectra for 10 gauge field configurations. 
    All spectra have been computed for anti-periodic fermionic boundary 
    conditions in the time direction.
  }
  \label{fig:manyspectra}
\end{figure}

Let us first look at the plots in Figure~\ref{fig:manyspectra}. 
On the left hand side, we show a superposition of ten spectra 
for the original (``raw'') configurations. 
The eigenvalues are close to the Ginsparg-Wilson circle, just as one 
would expect for the Chirally Improved Dirac operator. 
The eigenvalues extend all the way down to real eigenvalues, which 
shows that the underlying configurations are in the chirally broken 
phase.  There are also many real eigenvalues, i.e.,\ there are 
Dirac zero modes connected with the raw configurations. This shows that
many raw configurations are topologically non-trivial. 

The spectra of the vortex removed configurations show a completely 
different picture: There are no more real eigenvalues and thus 
no more topological modes. In addition, a large gap opened around 
the origin. The Banks-Casher relation \cite{Banks:docm} 
tells us, that we are now in the chirally restored phase. 
Since the eigenvalues stay close to the Ginsparg-Wilson circle, 
we know that our Dirac operator works well for vortex removed 
configurations. 

A prominent feature is, that the eigenvalues for vortex removed 
configurations form clusters in the complex plane. If one looks 
closely, 
one sees that there are eight eigenvalues per configuration in 
the lowest cluster. We would like to point out that the degeneracy 
of the lowest eigenvalue for a free field is also eight 
(anti-periodic temporal b.c.).
The spectrum of the free field is shown in the 
right hand side plot of Figure~\ref{fig:manyspectra}. 

\solSetCommonWidth{0.5\textwidth}{\ignorespaces
\includegraphics{eigen_plot_B250F04ph05_normal_forconf}\ignorespaces
\includegraphics{eigen_plot_B250F04ph05_removed_forconf}\ignorespaces
\includegraphics{eigen_plot_B250F04ph05_free_forconf}}

\section{Distribution of Eigenvalues}

We want to probe the clustering of eigenvalues, as mentioned in the 
previous section. In particular, we want to know if there is a 
connection between the number of eigenvalues in a cluster and the 
degeneracy of the eigenvalues for a trivial field. 
For this purpose  we modify the boundary condition of the fermion field in
the time direction: When the fermion fields wind once around the 
lattice in the time direction, they can acquire a phase. All other 
directions have periodic boundary conditions. Thus the eigenvectors 
$\vec v$ obey
\begin{gather}
  \label{eq:boundarycondition}
  \vec v(x+L\,\hat\imath\;)=\vec v(x),\quad
  \hat\imath=\widehat 1,\,\widehat 2,\,\widehat 3,\qquad
  \vec v(x+L\,\widehat 4\;)=e^{i2\pi\zeta}\,\vec v(x),
\end{gather}
where $L$ is the size of the lattice, $\hat\imath$ are the unit 
vectors in the spatial directions and $\widehat 4$ is the unit 
vector in the time direction. 

The important fact is, that the degeneracy of the eigenvalues 
for a free field depends on the parameter $\zeta$. 
It is $8$ in the case of \mbox{$\zeta=0$}, \mbox{$\zeta=1/2$}
and it is $4$ in the case of \mbox{$\zeta=1/4$}. 

\begin{figure}[t!]
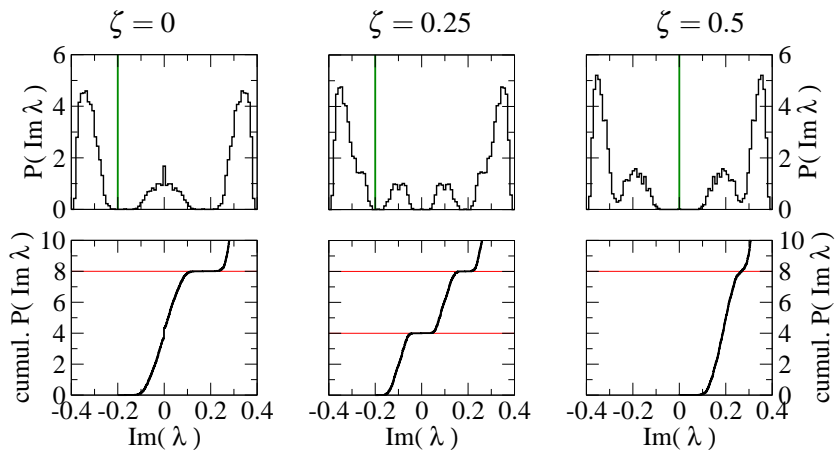

   \solSetCommonWidth{0.7\textwidth-6\arraycolsep}
  {\ignorespaces
    \includegraphics[clip]
    {hist-checkDegeneracy-fullvertical-vrm-ph00-B250_fortex}\ignorespaces
    \includegraphics[clip]
    {hist-checkDegeneracy-fullvertical-vrm-ph25-B250_fortex}\ignorespaces
    \includegraphics[clip]
    {hist-checkDegeneracy-fullvertical-vrm-ph50-B250_fortex}\ignorespaces
  }

  \begin{center}
    \begin{tabular}{@{}ccc@{}}
      \mbox{$ \zeta=0 $}     & 
      \mbox{$ \zeta=0.25 $}  & 
      \mbox{$ \zeta=0.5  $}  \\
      \includegraphics[clip,keepaspectratio,scale=\solGetCommonScale]
      {hist-checkDegeneracy-fullvertical-vrm-ph00-B250_fortex} &
      \includegraphics[clip,keepaspectratio,scale=\solGetCommonScale]
      {hist-checkDegeneracy-fullvertical-vrm-ph25-B250_fortex} &
      \includegraphics[clip,keepaspectratio,scale=\solGetCommonScale]
      {hist-checkDegeneracy-fullvertical-vrm-ph50-B250_fortex} 
    \end{tabular}
  \end{center}

  \caption[Cumulative Histograms]
  { Histograms and cumulative histograms for the 
    imaginary part of the Dirac spectra.
  }

  \label{fig:cumulativehistograms}
\end{figure}

Figure~\ref{fig:cumulativehistograms} 
shows histograms and cumulative histograms of Dirac spectra for 
vortex removed configurations. 
We plot the density of eigenvalues as a function of
\mbox{$\text{Im }\lambda$}. 
The upper row shows regular, normalized histograms. 
The lower row shows cumulative histograms, normalized by the 
number of configurations used.
The vertical line marks the starting point for the cumulative histograms. 
Clearly, we see that the lumpy structure of the spectrum as seen 
in Figure~\ref{fig:manyspectra} is reflected by the peaks in the 
regular histograms. These peaks then show up as plateaus and kinks 
in the cumulative histograms.  The horizontal lines mark 
the plateaus and kinks. We see that the position of plateaus and kinks 
of the cumulative histograms nicely matches the degeneracy of the 
free Dirac operator for a given value of $\zeta$. 
This is an indication that the vortex removed configurations 
are almost trivial configurations.

\section{The Effects of Random Changes}

The removal of vortices is quite a drastic change to the configuration. 
In our case, almost half of the links were changed. However, these 
changes occur in a highly non-trivial manner. 
Yet, one could criticize the vortex removal: Suppose the underlying 
topological structures of the gauge field are sensitive to arbitrary 
small changes of the configuration. This would mean that vortex removal 
cannot make a statement about the nature of the topological structures. 
To invalidate this critique, we apply random changes to the 
configuration. 
If the topological structures survive this test, we conclude that the 
vortices are strongly correlated with the relevant topological 
structures. 

One question remains: How many links do we have to change at random 
in order to perform a sensible test? The number of links, that are 
changed to remove the vortices, is surely not a good quantity. 
Note that the gauge fixing in Eq.~\eqref{eq:gaugefixing} does not 
fix the gauge completely: There is a 
remaining \mbox{$\mathbb{Z}_2$} gauge freedom that can be used to 
vary the number of negative links.
Thus, instead of using the total number of random changes, 
we use the average plaquette \mbox{$\left<U_p\right>$} 
as a measure to determine how many links we should change randomly. 
Table~\ref{tab:changedlinks} gives the overview of the situation.
For our randomly changed configurations, we changed 
1000~links, which corresponds to changing \mbox{$1.2\%$} 
of the links on a \mbox{$12^4$} lattice. 

\begin{table}
  \begin{center}
    \begin{tabular}{l|>{\ttfamily}r|>{\ttfamily}r|>{\ttfamily}r|} 
      ~ &
      {\rmfamily links changed} & 
      {\rmfamily plaquettes changed} 
      & \mbox{$ \left<U_p\right> $}\\ \hline
      original (raw) configurations   &  none    &   none  &  0.651 \\ \hline
      vortex removed configurations   &  50.0\%  &   3.3\% &  0.621 \\ \hline
      randomly changed configurations  &  1.2\%
                                      &  4.6\% 
                                      &  0.590 \\ 
                                      &   6.0\%  &  20.1\% &  0.389 \\
                                      &  24.1\%  &  46.5\% &  0.046 
    \end{tabular}
  \end{center}

  \caption[number of changed links]
  { Comparison of the number of changed links, changed plaquettes and 
    the change of the average plaquette for vortex removed and 
    randomly changed configurations. 
  }

  \label{tab:changedlinks}
\end{table}

\begin{figure}[t!]
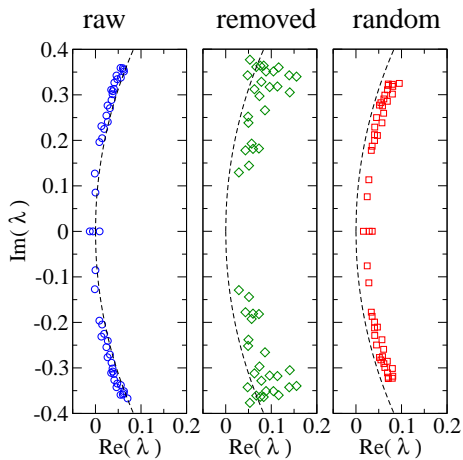

  \solSetCommonWidth{0.4\textwidth}{\ignorespaces
    \includegraphics{compare_print_pseudowoc-B250F04-002-ph05-forconf-show_raw}\ignorespaces
    \includegraphics{compare_print_pseudowoc-B250F04-002-ph05-forconf-show_woc}\ignorespaces
    \includegraphics{compare_print_pseudowoc-B250F04-002-ph05-forconf-show_pwc}}
        
  \begin{center}
    \begin{tabular}{@{}c@{}c@{}c@{}}
      raw & removed & random \\ 
      \includegraphics[clip,scale=\solGetCommonScale]
      {compare_print_pseudowoc-B250F04-002-ph05-forconf-show_raw}
      &
      \includegraphics[clip,scale=\solGetCommonScale]
      {compare_print_pseudowoc-B250F04-002-ph05-forconf-show_woc}
      &
      \includegraphics[clip,scale=\solGetCommonScale]
      {compare_print_pseudowoc-B250F04-002-ph05-forconf-show_pwc}
    \end{tabular}
  \end{center}

  \caption[single spectrum]
  { Dirac spectrum of a single configuration.
    It is shown for the unchanged, (raw) configuration, the same 
    configuration with the vortices removed, and with 
    random changes applied to the raw configuration. 
  }

  \label{fig:singlespectrum}
\end{figure}

\begin{figure}[t!]
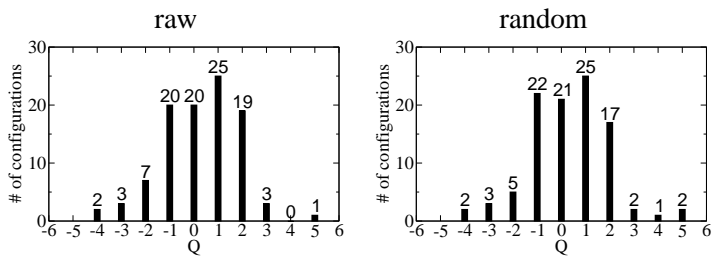

  \solSetCommonWidth{0.6\textwidth-2\arraycolsep}
  {\includegraphics{hist-topcharge-B250-04F-ph05_forconf}\ignorespaces
   \includegraphics{hist-topcharge-pwc1-B250-04F-ph05-D1b250vrm_coeffs_forconf}}
  
  \begin{center}
    \begin{tabular}{@{}cc@{}}
      raw & random \\
      \includegraphics[clip,scale=\solGetCommonScale]
      {hist-topcharge-B250-04F-ph05_forconf}
      &
      \includegraphics[clip,scale=\solGetCommonScale]
      {hist-topcharge-pwc1-B250-04F-ph05-D1b250vrm_coeffs_forconf}
    \end{tabular}
  \end{center} 

  \caption[topological charge]
  { This figure shows the distribution of the topological charge 
    $Q$ for raw and randomly changed configurations. 
    Vortex removed configurations always have $Q=0$. 
  }
  
  \label{fig:topcharge}
\end{figure}

Figure~\ref{fig:singlespectrum} shows the effect for a single 
configuration. As discussed above, the vortex removed spectrum looks 
completely different. In contrast, the spectrum for the randomly 
changed configuration looks essentially like the original (raw) 
spectrum. In particular, the number of zero modes did not change by the
random changes.
In Figure~\ref{fig:topcharge} we show the distribution of the topological 
charge for 100 configurations before and after random changes. 
We see that the overall shape of the distribution did not change. 

\section{Conclusions and Outlook} 

We can conclude that removal of the vortices produces configurations that 
resemble free fields. These vortex removed fields are 
void of topological structures and chiral symmetry is restored. 
Applying random changes to the original configuration shows that 
the removal of vortices is indeed a sensible method to probe the 
connection between vortex-like structures and topology. 
For \mbox{$\mathbb{Z}_2$} projected configurations the situation is 
different. In the case of \mbox{$\mathbb{Z}_2$} configurations that 
have been derived from thermalized configurations, the fermionic 
analysis does not yield sensible results. However, in the case of 
hand constructed vortices, analytic results are reproduced. 
Especially, the low-lying, non-zero modes of the Dirac operator do feel 
the presence of vortices, even for center projected configurations. 
For a detailed discussion of hand constructed configurations 
see~\cite{Gattnar:2004gx}. 

\bibliography{docm}


\end{document}